\documentclass[natbib209]{iaus}
\usepackage{graphicx}

\newcommand{\msun}{M$_\odot$}
\newcommand{\rsun}{R$_\odot$}
\newcommand{\lsun}{L$_\odot$}
\newcommand{\kms}{km~s$^{-1}$}

\title[Binary Central Stars] 
{Binary Central Stars of Planetary Nebula}

\author[O. De Marco]   
{Orsola De Marco}%

\affiliation{Department of Astrophysics, American Museum of Natural History, \\
Central Park West at 79th Street, New York, NY 10024, USA \break email: orsola@amnh.org\\[\affilskip]
 }

\pubyear{2006}
\volume{234}  
\pagerange{}
\date{?? and in revised form ??}
\jname{Planetary Nebulae in our Galaxy and Beyond}
\editors{M. J. Barlow \& R. H. M\'endez, eds.}
\begin{document}

\maketitle

\begin{abstract}
Only a handful of binary central stars of planetary nebulae (PNe) are known today, due to the difficulty of detecting their companions. Preliminary results from radial velocity surveys, however, seem to indicate that binarity plays a fundamental, rather than marginal role in the evolution of PNe and that the close binary fraction might be much larger than the currently known value of 10-15\%. In this review, we list all the known binary central stars, giving an updated census of their numbers and selected characteristics. A review is also given of the techniques used to detect binaries as well as selected characteristics of related stellar classes which might provide constraints (or additional puzzles) to the theory of PN evolution. Finally, we will formulate the conjecture that all PNe derive from binary interactions and suggest that this is not inconsistent with our current knowledge.

\end{abstract}

\firstsection 
\section{Introduction}
\label{sec:introduction}
For the last three decades astronomers in our field have argued over what causes the variety of planetary nebula (PN) morphologies. The bone of contention has been whether a single star is capable of generating axi-symmetric and point-symmetric PNe. Single star models, including stellar rotation and magnetic fields (e.g., Garc{\'{\i}}a-Segura et al. 2005)\nocite{Gar+05} have succeeded in reproducing highly collimated PN shapes. It is argued, however (Soker 2006)\nocite{Sok06}, that {\it global} magnetic fields cannot survive in the star for long, since they would quickly slow the star down. A companion is therefore needed to supply angular momentum to the star and sustain the rotation and magnetic field. Single star models  do not couple the magnetic field to the stellar envelope, thus creating an unphysical situation. 

Binary models  have successfully reproduced PN morphologies (e.g., Garc\'ia-Arredondo \& Frank 2004)\nocite{GF04} but this by itself does not prove that binaries are necessary, only that they are sufficient. From a theoretical point of view, we can therefore {\it suspect} that a companion might be needed to create PNe with axi-symmetric or point-symmetric shapes, but we cannot be certain. Also, we do not know what type of companion and what type of interaction might be needed.

It is therefore likely that theory does not hold the answer to the problem of PN binarity. The answer must therefore come from observations of the actual central star binary fraction, period distribution and companion mass distribution. This however, is no simple problem. First of all, the high intrinsic luminosity of central stars and their relatively large radii (compared to those of bona fide white dwarfs [WDs]) make them very bright even in the near infrared. This, combined with their relatively large distances ($>$0.5~kpc), makes it very difficult to detect late type companions by searching for excess infrared flux in central star spectra. Second, the radial velocity (RV) technique used to detect orbiting companions works most effectively on bright stars with many thin absorption lines. Central stars of PNe tend to be faint (because they are far) and often have few and broad stellar absorption lines, contaminated by the PN hydrogen and helium emission lines.
Despite this, the attention that the central star binarity problem is getting, makes one confident that substantial headway will be made before the next PN meeting in 2010.
  
\section{Detection techniques and binary central star statistics}
\label{sec:detection}

The bulk of the short period central star binaries was discovered by looking for periodic photometric variability in a sample of 100 central stars (Bond 2000)\nocite{Bon00}. Thirteen were found to vary with period shorter than 3 days. As a result of this survey, the central star binary fraction is today thought to be 10-15\%.  The photometric technique relies on heating of the cool companion by the hot central star, or on ellipsoidal variability due to the companion's distortion because of the primary's gravity. This technique is therefore invariably sensitive only to very short orbital separations. It is  therefore reasonable to conclude that 10-15\% is a lower limit to the true central star binary fraction.

To detect longer period binaries, one has to resort to  RV surveys. In Table~\ref{tab:RVsurveys} we list the four RV surveys from the literature. RV (spectroscopic) surveys are much harder to carry out than photometric ones, mostly because relatively long observing runs, spanning up to months, need to be carried out on moderate size telescopes (4-m or above) with good
spectral resolutions. 

 \cite{Men89} took one or two high resolution spectrograms of each central star in a sample of 28 objects. When  two spectra were available, spectral line wavelength shifts were searched for, when not, stellar line wavelengths were compared to those of PN lines. The conclusion from that survey was that none of the objects was variable beyond doubt, in particular since several objects might be wind variables (where the lines change in shape and position due to inhomogeneities in the stellar wind).  

More recently,
\cite{SP03} carried out a survey at intermediate resolution, and determined that 39\% of their sample were RV variables. By assessing the various biases in their survey (a bias {\it for} binaries because late type companions would make the central star brighter and more easily observed, and a bias {\it against} binaries because of inclination and mass effects), they argued that the true central star close binary fraction should be $\sim$60\%. The real caveat of their survey, however, is that no periods were detected so that wind variability is still an alternative explanation for the presence of RV variability in several objects.

Finally, De Marco et al. (2004)\nocite{DeM+04} and \cite{AB05} teamed up to carry out a survey of both hemispheres. They detected 91\% and 37-50\% RV variable fractions, respectively (although their sample sizes were undoubtedly small). The smaller RV variable fraction of the \cite{AB05} survey is due to the lower resolution of their setup ($\sim$10~\kms\ instead of $\sim$3~\kms), which is also similar to the resolution used by \cite{SP03}.  

During a recent observing run at the Kitt Peak 4-m telescope, we (Bond \& De Marco) have monitored two (IC4593 and BD+33~2642) of the RV variables of De Marco et al. (2004)\nocite{DeM+04} to look for periods using an echelle setup. Preliminary results show that both these central stars have periods between 4 and 5 days. Especially interesting is the fact that one of the two stars, IC4593, is indeed a wind variable as was previously suspected (its HeII $\lambda$4686 line has a variable P~Cygni profile), but when weak, optically thin emission lines are analyzed, a periodicity is revealed. The other star, BD+33~2642, is a flux standard and it is therefore surprised that  its RV variability was not detected previously. This star, is also peculiar because it is a post-asymptotic giant branch (AGB) star with an effective temperature of 20\,200~K (Napiwotzki 1999)\nocite{Nap+94}. It might therefore not be hot enough to ionize its PN, and could be the companion of an invisible pre-WD central star. Assuming a maximum central star luminosity of 15\,000~\lsun, its maximum possible radius would be 10~\rsun. The minimum possible orbital period would therefore be 3.4--4.2~days (for a 0.6-\msun\ primary and a 0.6--0.2-\msun\ companion). This is consistent with the detected period.

Aside from the photometric and RV surveys, a handful of binary central stars have also been detected because of their composite spectra, where a hot star, detected in the UV  has a cooler component detected in the optical regime.  

For completeness, we should also mention that 9-14\% of 113 surveyed targets appear to have distant companions ($a >> 100$~AU). These companions are unlikely to have affected the evolution of the central star. It is unlikely that they even participated in the shaping of the PN by the most feeble mechanism (focusing; Gawryszczak et. al 2002)\nocite{Gaw+02}. It is not excluded that the central star primary in these wide pairs might have additional close companions which are responsible for shaping the PN. This, for instance, could be the case for A63, which was classified as a possible wide binary by Ciardullo et al. (1999)\nocite{Cia+99}, with a secondary at 3440~AU, but is also known to be an eclipsing binary (Table~\ref{tab:BCS}). 

\begin{table}
\begin{tabular}{lcccccc}
RV survey         & Res.  &Telescope &	\# 	&\#          & SNR&	\% RV  	  \\
                         & (\AA) &                 &  Obj.   &  Meas. &        &Variables \\
\hline
\cite{Men89}	& 0.3	          & 3.6-m ESO  	&28	& 1-2   & 	--        &      0\%          \\
\cite{SP03}	& 1.5                   & 2.5-m INT	         & 33	& 6-40 & ~100   &	39\%        \\  
De Marco et al. (2004)\nocite{DeM+04}	& 0.6  & 3.6-m WIYN 	& 11 & 6-16 & ~30-50&	91\%         \\
 \cite{AB05}	& 1.5  & 1.5-m CTIO        & 19 &  5-47 & ~30-50& 	37-50\%  \\
 \hline
\end{tabular}
\caption{RV surveys of central stars of PNe. The brightness limit of these surveys is $\sim$14-15~mag. For an explanation of the acronyms used, see text.}
\label{tab:RVsurveys}
\end{table}

 \section{Known close binary central stars}\label{sec:BCS}

 In Table~\ref{tab:BCS} we present the list of PNe with confirmed binary central stars. We consider `central stars' only stars that have gone through the AGB evolution. Although we think that it would be best to stick to classifying PNe by observed, rather than interpretative characteristics, a certified post-AGB nature of the primary appears to be the community's consensus for the inclusion in the central star class. So, for instance, the star EGB5, a photometric binary with period 1.18~day (or, less likely, 0.55~day; Mendez \& Niemela 1981; Karl et al. 2003)\nocite{MN81,Kar+03}, in the middle of a nebulosity, is classified as a post-red giant branch (RGB) star by a comparison with evolutionary calculations, and is therefore excluded from Table~\ref{tab:BCS}. We also point out that central stars are almost never WDs, but rather {\it pre}-WDs, since their radii are almost always about ten times larger than those of regular WDs (0.1-1.0~\rsun; Napiwotzki 1999\nocite{Nap99}). We also warn against calling central stars `subdwarfs', since this introduces confusion with post-RGB objects (i.e., the sdB stars, which are hot horizzontal branch stars). Finally, we also want to stress that there is a fuzzy line between post-AGB stars and central stars of PN. In theory, post-AGB stars are not yet hot enough to ionize the PN and if they have a nebulosity around them, this is detected in reflected stellar light. There are, however, central stars of ionized PNe that have temperatures thought too low to do the ionizing. In these cases, it is thought that the ionizing source is an invisible WD companion.  
 
Returning to Table~\ref{tab:BCS}, the PN designation (column 1) is followed, in column 2, by the coordinates from
Kerber et al. (2003)\nocite{Ker+03}, expressed to a precision of one tenth of a second of time and one second of arc. In Column 3 we list the $V$ magnitudes of the central star. These are only meant as indicative numbers,  since in most cases the central stars vary in brightness (the source of these values are the Simbad Database, the ESO PN catalogue [Acker et al. 1992]\nocite{Ack+92} or Tylenda et al. [1989]\nocite{Tyl+89}). In column 4 we indicate the type of binary: photometric binaries (P) are discovered by periodic light variability due to either an illumination effect, or to ellipsoidal variations; spectroscopic binaries (S) have {\it periodic} radial velocity shifts of the central star lines (all spectroscopic central star binaries are single lined spectroscopic binaries); eclipsing binaries (E) have periodic light variations due to the secondary eclipsing the primary; composite binaries (C) are stars whose spectra betray the presence of a hot component, shining brightly in the UV and a cooler one shining brightly in the red or near-infrared part of the spectrum. In column 5 we indicate the binary period (truncated to three decimal places), followed, in column 6, by the spectral type of the companion.  `Cool' means that the companion is likely a cool main sequence star, (more likely of spectral type M rather than K) and not, for instance, a WD. Finally, in column 7, we list the references.
 
 \subsection{Additional characteristics of selected binary central stars}
 \label{ssec:individual}
  
HFG1 and A65. \cite{AS90} and \cite{WW96} report these central stars to be similar to intermediate polar cataclysmic variables (AM Her stars) and to have an accretion disk around the hot component.  
 
NGC1514. This central star displays a composite spectrum (Feibelman 1997)\nocite{Fei97}. However, neither photometric, nor spectroscopic variability has been detected (Bond \& Grauer 1987)\nocite{BG87}, nor has this star a distant companion (Ciardullo et al. 1999)\nocite{Cia+99}. The period thus remains undetermined.
 
A35, LoTr1 and LoTr5. These PNe harbor binary central stars discovered because of their composite spectra. The two components of the binaries are thought to be at intermediate separations and the companions are thought to be evolved. If this is the case, these would be rare systems indeed since detecting a binary where both stars are in short-lived evolutionary phases (giant and central star of PN, for the secondary and primary, respectively) should require extreme fine tuning of the masses. It is therefore surprising that we know three such systems, unless their brightness makes them conspicuous. All three systems exhibit periodic variations (A35: 0.76 or 3.3-day; Bond et al. 1993\nocite{Bon+93}; LoTr1: 6.6-day; Bond et al. 1993\nocite{Bon+93}; LoTr5: 5.9 day; Strassmeier et al. 1997\nocite{Str+97}) attributed to star spots on the surface of the fast rotating (for the spectral type) companions. This is thought to be due to accretion of material with high specific angular momentum either during a common envelope phase or some other type of strong interaction. None of the three binaries exhibits RV variability (A35, LoTr1: \cite{AB05} to a precision of $\sim$10~\kms; LoTr1:  Strassmeier et al. 1997). Gatti et al. (1998)\nocite{Gat+98} determined the separation of the two components in A35 to be between 0.08'' and 0.14'', which, at the adopted distance of 160~pc, translates into a separation of 13-28~AU. The central star radius was determined to be 0.1~\rsun (Herald \& Bianchi 2002)\nocite{HB02}.  LoTr5 has been reported to be a triple system, where the G5III star orbits at an intermediate separation around a close binary. However, there is no conclusive evidence for any periodicity in the RV data (see Strassmeier et al. (1997)\nocite{Str+97} for a history of the periodicity of this system).
     
NGC6826.  \cite{Nos80} and \cite{Ack+82} suggested that this star is a binary and it is listed as a spectroscopic binary by \cite{Men89}. It is listed as a confirmed binary in Table~\ref{tab:BCS}, but the two publications mentioned above could not be retreived and checked.

 \begin{table}\def~{\hphantom{0}}
  \begin{center}
  \caption{Known close binary central stars. P: photometric; S: spectroscopic; E: eclipsing; C: composite spectrum.}
  \label{tab:BCS}
  \begin{tabular}{lllllll}\hline
      PN name  & RA \& Dec   &   V mag & Binary   & Period  &  Comp. & Reference \\ 
         &     &     &   Class & (days)    &Sp. Type   &   \\\hline

HFG1	&	03 03 47.0 +64 54 35	&	13.4  	& P, S	&	0.582        &   cool       	&	Grauer et al. (1987)\nocite{Gra+87} \\
&&&&&& \cite{AS90}	\\
NGC1360	&	03 33 14.6 --25 52 18	&	11.2	         &	S	&	8.2		&  WD/MS	&	\cite{MN77}\\
 &&&&&&Hoare et al. (1996)\nocite{Hoa+96}	\\
NGC1514	&	04 09 17.0 +30 46	         &	9.4	         &       C	&	--       	&   A0-3III 	&       \cite{Fei97} \\
LoTr1	&	05 55 06.7 --22 54 02	&	12.5	         &	C	&	--	         &   KV           	&	Bond et al. (1989)\nocite{Bon+89}	\\
NGC2346	&	07 09 22.5 --00 48 24	&	13.8B	&	S	&	15.99	&   A0III	         &	\cite{MN81}	\\
K1-2	         &	08 57 45.9 --28 57 36	&	16.6	        &	P	&	0.676	      & MV	         &       Exter et al. (2003)\nocite{Ext+03}		\\
DS1	         &	10 54 40.6 --48 47 03 	&	12.3  	&	P	&	0.357              & MV	         &	Kilkenny et al. (1988)\nocite{Kil+88}\\
                   &&&&&& \cite{Bon00}	\\
PNG136$^a$&11 53 24.7 +59 39 57     &	17.9	        &	S, P  &	0.163	      &WD?	&	Tovmassian et al. (2004)\nocite{Tov04}\\
                   &&&&&& Jacoby et al. (these proceedings)	\\
 BE UMa   &	11 57 44.8 +48 56 19	&	14.8	        &	E	&	2.29	         &M4V&	Ferguson et al. (1999)\nocite{Fer+99}	\\
                    &&&&&&  Wood et al. (1995)\nocite{Woo+95}	\\
A35	         &	12 53 32.8 --22 52 23	&	9.6           &	C	&      $>$10yr	& G8III-IV	&	Gatti et al. (1997, 1998)\nocite{Gat+97,Gat+98}\\
                   &&&&&&  \cite{HB02}	\\         
LoTr5	&	12 55 33.7 +25 53 31	&	8.8	       &	C	&	--	         	      &G5III	&	\cite{FK83}	\\
SuWt2	&	13 55 43.2 --59 22 40	&	12.3	       &	E	&	2.45	         	      &B9V	&	\cite{Wes76,Bon00}	\\
Sp1    	&	15 51 40.9 --51 31 28	&	14.0	&	P,C?	&	2.91	         		& cool	&	\cite{BL90}\\
                   &&&&&& M\'endez et al. (1988)\nocite{Men+88}	\\
NGC6026	&	16 01 21.1 --34 32 37	&	13.3	&	P, S	&	0.528			& M0V      &	Hillwig et al. (these proceedings)	\\
HaTr4	&	16 45 00.2 --51 12 20	&	17.1	&	P	&	1.74	         		&   ?    &	\cite{BL90}	\\
NGC6302	&	17 13 44.2 --37 06 16	&	16.0	&	C 	&	--           &GV	&	\cite{Fei01}	\\
 A41	         &	17 29 02.0 --15 13 04	&	16.5	&	P	&	0.113	 &GV	&	\cite{GB83}	\\
NGC6337	&	17 22 15.7 --38 29 03	&	14.9	&	P	&	0.173		&M4V	&Hillwig et al. (these proceedings)		\\
A46	         &	18 31 18.3 +26 56 13	&	15.1	&	E	&	0.472		&	M6V	&	\cite{PB94}	\\
Hf2-2	&	18 32 30.9 --28 43 20	&	18.0B	&	P	&	0.399			&cool		&	\cite{Bon00}	\\
Sh2-71	&	19 01 59.3  +02 09 18	&	13.8 &	C?P?&	17-22	&A7V-F0V&	\cite{Koh79}\\
                   &&&&0.5-1&& \cite{Fei99}	\\
 A63	          &	19 42 10.4 +17 05 14	&	15.1	&	E	&       0.456	  & M2V-M5V	&	Bond et al. (1978)\nocite{Bon+78}\\
                   &&&&&& \cite{PB93}	\\
NGC6826	&	19 44 48.1 +50 31 30	&	10.1	&	S       &	0.238			&	?	&	\cite{Men89}	\\
A65	          &	19 46 34.2 --23 08 13	&	15.9	&	E	&	$\sim$1	         		&	cool	& \cite{WW96}		\\
HB12	&	23 26 14.8 +58 10 55	&	13.8	&	E	&	0.141	& cool	         &	Hsia et al. (2006)\nocite{Hsi+06}\\
       \hline
  \end{tabular}
  $^a$PN G135.9+55.9
 \end{center}
\end{table}

\section{Related classes: post-AGB stars and WDs}

Post-AGB stars are supergiants with spectral types between G and B. They have left the AGB and are shrinking and heating on their way, it is thought, to becoming central stars of ionized PNe. They can be surrounded by envelopes visible in reflected stellar light (they are not yet ionized, although in some cases ionization is setting in, e.g., in the case of LSIV-12 111 [Conlon et al. 1993]\nocite{Con+93}). Post-AGB star nebulae often have extremely bipolar morphologies within more spherical envelopes (Sahai \& Trauger 1998)\nocite{ST98}.

The post-AGB star binary fraction, period and companion mass distributions would be a valuable piece of information in understanding the role of binarity in making PNe. This fraction is alas not known. A curious fact is, however, reported in the literature.
Van Winckel (2003)\nocite{van03} reports a very large binary fraction ($\sim$100\%) in post-AGB stars {\it previously known to have large dusty disks}. The binary periods are in the range 100-1500~days. The companions are thought to be unevolved. The companion masses are not known but are likely to span a wide range. Strangely, the orbits' eccentricities are zero for periods smaller than 300 days, but much higher for longer periods. These objects must have undergone strong interaction between the primary and the companion, since the AGB precursor's radius is likely to have extended beyond the present day orbit of the companion. It is a mystery how these systems have avoided the common envelope phase. 

One question to ask is what is the relationship between these systems and PNe. If these post-AGB stars do eventually ionize their PNe, central star binaries in this period range should exist, although they would be admittedly hard to detect.  They could be the progenitors of A35 type binary central stars. It has also been suggested that they are the progenitors of symbiotic binaries (van Winckel 2003)\nocite{van03}. 


Between 22 and 33\% of the WD population is thought to be in binary systems with cooler, unevolved companions or other WDs (Green et al. 2000; Farihi et al. 2005)\nocite{Gre+00,Far+05}. Further, Farihi et al. (2006)\nocite{Far+06} determined that $\sim$66\% of a sample of binary WDs are totally or partly resolved (by the {\it Hubble Space Telescope}  {\it Advanced Camera for Surveys}) into binary and triple systems, while the rest consist of unresolved binaries. The binaries have a bimodal period distribution with a gap at $\sim$1~AU. This is expected, in view of the fact that systems that enter a common envelope either on the RGB or the AGB will result in closer binaries (or mergers) and only systems with separations in excess of $\sim$2-3~AU (or up to a few times more, if tidal capture operates) will be left alone. 

The spectral types of the companions for the Farihi et al. (2006)\nocite{Far+06} sample peak at around M4 (both for resolved and unresolved binaries). The latest companion type in the unresolved sample is M7, possibly suggesting (although we are in the limit of low number statistics, with only 13 systems) that the smallest mass that can emerge from a common envelope interaction is 0.1~\msun (interpolating between the masses of M5 and M8 spectral types  (Cox 2000)\nocite{Cox00}). This is in agreement with the calculations of De Marco et al. (2003)\nocite{DeM+03}.

If one-third of all WDs are binaries and one third of those are close binaries, then $\sim$10\% of WDs are in post-common envelope {\it close} binaries. If all WDs have gone through a PN phase, then we expect that, just as is the case for WDs, only 10\% of central stars of PN should be in close binary systems. Conversely, if we start from the assumption that {\it all} central stars are in close binary systems, then one should conclude that  single WDs and WDs in wide binaries were never surrounded by a PN. In such case, we would expect the close binary WDs to be the descendants of the central stars of PN and the single and wide WD binaries to be the descendants of stars that  never made a visible PN. If this were the case, the PN birthrate density should be about one tenth of that of WDs (1.0$\times$10$^{-12}$~WD~pc$^{-3}$~yr$^{-1}$; Liebert et al. 2005)\nocite{Lie+05} and at odds with what was determined by Phillips (2002; 2.1$\times$10$^{-12}$~PN~pc$^{-3}$~yr$^{-1}$)\nocite{Phi02}, but more in line with the calculations of Moe \& De~Marco (these proceedings) .

\section{Discussion}

One question that does arise from the RV survey work is whether it is conceivable that PNe are purely or almost purely an interactive binary phenomenon. Although a final answer to this question must come from the determination of the  binary fraction, such a supposition can be checked for consistency with what we already know of stellar evolution and binarity.  

The binary fraction in PNe central stars should be the same as that of G-F main sequence stars ($\sim$60\%; Duquennoy \& Mayor 1991), or slightly higher, if we include slightly more massive stars. The central star period distribution is expected to have a gap since systems with separations  $<$500--1500~\rsun\ will have their orbits altered by interaction effects (including common envelopes). The central star {\it close} binary fraction should be $\sim$15\%, the same as  for main sequence binaries with separations $<$500-1500~\rsun\ (some systems will  merge reducing the binary fraction further). Contrary to this prediction,  RV surveys seem to point to a {\it close} binary fraction of the order of 60-90\%.  If this result is confirmed, one might conclude that binarity is required for the ejection of a PN and that single stars and wide binaries do not make a visible PN (Subag \& Soker 2005)\nocite{SS05}. If so, the PN population would be only a fraction of the population of stars transiting between the AGB and the WD domain.

Moe \& De Marco (2006 and in these proceedings)\nocite{MD06} tested this prediction by determining theoretically the number of Galactic PNe that descend from single stars and binaries and those deriving from post-common envelope binaries. The uncertainties of such predictions are large, as are those of the observational counterparts of these values, with which the predictions are compared. However, considering all the uncertainties the conclusion is that if PNe derive from single stars and binaries the predicted PN galactic population is significantly larger than is observed. If, on the other hand, only post-common envelope stars can make PNe, the predicted Galactic PN population is more in line with observations. Whether or not the statement above is correct, we can definitely conclude that a PN population dominated by short-period binaries is consistent with what we know of stellar evolution and the main sequence binary fraction and period distribution (Duquennoy \& Mayor 1991)\nocite{DM91}.  The final answer,
 however, rests in the success of the observational programs aimed at determining the central star binary fraction.

 \begin{acknowledgments}
 I owe a continuous stream of thanks to my collaborators Howard Bond and George Jacoby for teaching me the ropes of how to observe PNe and, although I am not Noam Soker, I am also indebted to him for initiating me to the many issues of binary central star research. This research has made use of the SIMBAD database,
operated at CDS, Strasbourg, France.
 
\end{acknowledgments}

\bibliographystyle{apj}                       

\begin{thebibliography}{61}
\expandafter\ifx\csname natexlab\endcsname\relax\def\natexlab#1{#1}\fi

\bibitem[{{Acker} \& {Gleizes}(1982)}]{Ack+82}
{Acker}, A. \& {Gleizes}, F. 1982, Publication Speciale du Centre de Donnees
  Stellaires, 3

\bibitem[{{Acker} {et~al.}(1992){Acker}, {Marcout}, {Ochsenbein}, {Stenholm},
  \& {Tylenda}}]{Ack+92}
{Acker}, A., {Marcout}, J., {Ochsenbein}, F., {Stenholm}, B., \& {Tylenda}, R.
  1992, {Strasbourg - ESO catalogue of galactic planetary nebulae. Part 1; Part
  2} (Garching: European Southern Observatory, 1992)

\bibitem[{{Acker} \& {Stenholm}(1990)}]{AS90}
{Acker}, A. \& {Stenholm}, B. 1990, A\&A, 233, L21

\bibitem[{{Af{\v s}ar} \& {Bond}(2005)}]{AB05}
{Af{\v s}ar}, M. \& {Bond}, H.~E. 2005, Memorie della Societa Astronomica
  Italiana, 76, 608

\bibitem[{{Bond}(2000)}]{Bon00}
{Bond}, H.~E. 2000, in ASP Conf. Ser. 199: Asymmetrical Planetary Nebulae II:
  From Origins to Microstructures, ed. J.~H. {Kastner}, N.~{Soker}, \&
  S.~{Rappaport}, 115--+

\bibitem[{{Bond} {et~al.}(1989){Bond}, {Ciardullo}, \& {Meakes}}]{Bon+89}
{Bond}, H.~E., {Ciardullo}, R., \& {Meakes}, M. 1989, BAAS, 21, 789

\bibitem[{{Bond} {et~al.}(1993){Bond}, {Ciardullo}, \& {Meakes}}]{Bon+93}
{Bond}, H.~E., {Ciardullo}, R., \& {Meakes}, M.~G. 1993, in IAU Symp. 155:
  Planetary Nebulae, ed. R.~{Weinberger} \& A.~{Acker}, 397--+

\bibitem[{{Bond} \& {Grauer}(1987)}]{BG87}
{Bond}, H.~E. \& {Grauer}, A.~D. 1987, in IAU Colloq. 95: Second Conference on
  Faint Blue Stars, ed. A.~G.~D. {Philip}, D.~S. {Hayes}, \& J.~W. {Liebert},
  221--228

\bibitem[{{Bond} {et~al.}(1978){Bond}, {Liller}, \& {Mannery}}]{Bon+78}
{Bond}, H.~E., {Liller}, W., \& {Mannery}, E.~J. 1978, ApJ, 223, 252

\bibitem[{{Bond} \& {Livio}(1990)}]{BL90}
{Bond}, H.~E. \& {Livio}, M. 1990, ApJ, 355, 568

\bibitem[{{Ciardullo} {et~al.}(1999){Ciardullo}, {Bond}, {Sipior}, {Fullton},
  {Zhang}, \& {Schaefer}}]{Cia+99}
{Ciardullo}, R., {Bond}, H.~E., {Sipior}, M.~S., {Fullton}, L.~K., {Zhang},
  C.-Y., \& {Schaefer}, K.~G. 1999, AJ, 118, 488

\bibitem[{{Conlon} {et~al.}(1993){Conlon}, {Dufton}, {McCausland}, \&
  {Keenan}}]{Con+93}
{Conlon}, E.~S., {Dufton}, P.~L., {McCausland}, R.~J.~H., \& {Keenan}, F.~P.
  1993, ApJ, 408, 593

\bibitem[{{Cox}(2000)}]{Cox00}
{Cox}, A.~N. 2000, {Allen's astrophysical quantities} (Allen's astrophysical
  quantities, 4th ed.~Publisher: New York: AIP Press; Springer, 2000.~Editedy
  by Arthur N.~Cox.~ ISBN: 0387987460)

\bibitem[{{De Marco} {et~al.}(2004){De Marco}, {Bond}, {Harmer}, \&
  {Fleming}}]{DeM+04}
{De Marco}, O., {Bond}, H.~E., {Harmer}, D., \& {Fleming}, A.~J. 2004, ApJ,
  602, L93

\bibitem[{{De Marco} {et~al.}(2003){De Marco}, {Sandquist}, {Mac Low},
  {Herwig}, \& {Taam}}]{DeM+03}
{De Marco}, O., {Sandquist}, E.~L., {Mac Low}, M.-M., {Herwig}, F., \& {Taam},
  R.~E. 2003, in Revista Mexicana de Astronomia y Astrofisica Conference
  Series, ed. M.~{Reyes-Ruiz} \& E.~{V{\'a}zquez-Semadeni}, 24--30

\bibitem[{{Duquennoy} \& {Mayor}(1991)}]{DM91}
{Duquennoy}, A. \& {Mayor}, M. 1991, A\&A, 248, 485

\bibitem[{{Exter} {et~al.}(2003){Exter}, {Pollacco}, \& {Bell}}]{Ext+03}
{Exter}, K.~M., {Pollacco}, D.~L., \& {Bell}, S.~A. 2003, MNRAS, 341, 1349

\bibitem[{{Farihi} {et~al.}(2005){Farihi}, {Becklin}, \& {Zuckerman}}]{Far+05}
{Farihi}, J., {Becklin}, E.~E., \& {Zuckerman}, B. 2005, ApJS, 161, 394

\bibitem[{{Farihi} {et~al.}(2006){Farihi}, {Hoard}, \& {Wachter}}]{Far+06}
{Farihi}, J., {Hoard}, D.~W., \& {Wachter}, S. 2006, ArXiv Astrophysics
  e-prints

\bibitem[{{Feibelman}(1997)}]{Fei97}
{Feibelman}, W.~A. 1997, PASP, 109, 659

\bibitem[{{Feibelman}(1999)}]{Fei99}
---. 1999, PASP, 111, 719

\bibitem[{{Feibelman}(2001)}]{Fei01}
---. 2001, ApJ, 550, 785

\bibitem[{{Feibelman} \& {Kaler}(1983)}]{FK83}
{Feibelman}, W.~A. \& {Kaler}, J.~B. 1983, ApJ, 269, 592

\bibitem[{{Ferguson} {et~al.}(1999){Ferguson}, {Liebert}, {Haas}, {Napiwotzki},
  \& {James}}]{Fer+99}
{Ferguson}, D.~H., {Liebert}, J., {Haas}, S., {Napiwotzki}, R., \& {James},
  T.~A. 1999, ApJ, 518, 866

\bibitem[{{Garc{\'{\i}}a-Arredondo} \& {Frank}(2004)}]{GF04}
{Garc{\'{\i}}a-Arredondo}, F. \& {Frank}, A. 2004, ApJ, 600, 992

\bibitem[{{Garc{\'{\i}}a-Segura} {et~al.}(2005){Garc{\'{\i}}a-Segura},
  {L{\'o}pez}, \& {Franco}}]{Gar+05}
{Garc{\'{\i}}a-Segura}, G., {L{\'o}pez}, J.~A., \& {Franco}, J. 2005, ApJ,
  618, 919

\bibitem[{{Gatti} {et~al.}(1997){Gatti}, {Drew}, {Lumsden}, {Marsh}, {Moran},
  \& {Stetson}}]{Gat+97}
{Gatti}, A.~A., {Drew}, J.~E., {Lumsden}, S., {Marsh}, T., {Moran}, C., \&
  {Stetson}, P. 1997, MNRAS, 291, 773

\bibitem[{{Gatti} {et~al.}(1998){Gatti}, {Drew}, {Oudmaijer}, {Marsh}, \&
  {Lynas-Gray}}]{Gat+98}
{Gatti}, A.~A., {Drew}, J.~E., {Oudmaijer}, R.~D., {Marsh}, T.~R., \&
  {Lynas-Gray}, A.~E. 1998, MNRAS, 301, L33

\bibitem[{{Gawryszczak} {et~al.}(2002){Gawryszczak}, {Miko{\l}ajewska}, \&
  {R{\'o}{\.z}yczka}}]{Gaw+02}
{Gawryszczak}, A.~J., {Miko{\l}ajewska}, J., \& {R{\'o}{\.z}yczka}, M. 2002,
  A\&A, 385, 205

\bibitem[{{Grauer} \& {Bond}(1983)}]{GB83}
{Grauer}, A.~D. \& {Bond}, H.~E. 1983, ApJ, 271, 259

\bibitem[{{Grauer} {et~al.}(1987){Grauer}, {Bond}, {Ciardullo}, \&
  {Fleming}}]{Gra+87}
{Grauer}, A.~D., {Bond}, H.~E., {Ciardullo}, R., \& {Fleming}, T.~A. 1987,
  BAAS, 19, 643

\bibitem[{{Green} {et~al.}(2000){Green}, {Ali}, \& {Napiwotzki}}]{Gre+00}
{Green}, P.~J., {Ali}, B., \& {Napiwotzki}, R. 2000, ApJ, 540, 992

\bibitem[{{Herald} \& {Bianchi}(2002)}]{HB02}
{Herald}, J.~E. \& {Bianchi}, L. 2002, ApJ, 580, 434

\bibitem[{{Hoare} {et~al.}(1996){Hoare}, {Drake}, {Werner}, \&
  {Dreizler}}]{Hoa+96}
{Hoare}, M.~G., {Drake}, J.~J., {Werner}, K., \& {Dreizler}, S. 1996, MNRAS,
  283, 830

\bibitem[{{Hsia} {et~al.}(2006){Hsia}, {Ip}, \& {Li}}]{Hsi+06}
{Hsia}, C.~H., {Ip}, W.~H., \& {Li}, R.~J.~Z. 2006, AJ, in press


\bibitem[{{Karl} {et~al.}(2003){Karl}, {Napiwotzki}, {Heber}, {Lisker},
  {Nelemans}, {Christlieb}, \& {Reimers}}]{Kar+03}
{Karl}, C., {Napiwotzki}, R., {Heber}, U., {Lisker}, T., {Nelemans}, G.,
  {Christlieb}, N., \& {Reimers}, D. 2003, in NATO ASIB Proc. 105: White
  Dwarfs, ed. D.~{de Martino}, R.~{Silvotti}, J.-E. {Solheim}, \& R.~{Kalytis},
  43--+

\bibitem[{{Kerber} {et~al.}(2003){Kerber}, {Mignani}, {Guglielmetti}, \&
  {Wicenec}}]{Ker+03}
{Kerber}, F., {Mignani}, R.~P., {Guglielmetti}, F., \& {Wicenec}, A. 2003,
  A\&A, 408, 1029

\bibitem[{{Kilkenny} {et~al.}(1988){Kilkenny}, {Spencer Jones}, \&
  {Marang}}]{Kil+88}
{Kilkenny}, D., {Spencer Jones}, J.~H., \& {Marang}, F. 1988, The Observatory,
  108, 88

\bibitem[{{Kohoutek}(1979)}]{Koh79}
{Kohoutek}, L. 1979, Informational Bulletin on Variable Stars, 1672, 1

\bibitem[{{Longmore} \& {Tritton}(1980)}]{LT80}
{Longmore}, A.~J. \& {Tritton}, S.~B. 1980, MNRAS, 193, 521

\bibitem[{{Mendez}(1989)}]{Men89}
{Mendez}, R.~H. 1989, in IAU Symp. 131: Planetary Nebulae, ed.
  S.~{Torres-Peimbert}, 261--272

\bibitem[{{Mendez} {et~al.}(1988){Mendez}, {Kudritzki}, {Herrero}, {Husfeld},
  \& {Groth}}]{Men+88}
{Mendez}, R.~H., {Kudritzki}, R.~P., {Herrero}, A., {Husfeld}, D., \& {Groth},
  H.~G. 1988, A\&A, 190, 113

\bibitem[{{Mendez} \& {Niemela}(1977)}]{MN77}
{Mendez}, R.~H. \& {Niemela}, V.~S. 1977, MNRAS, 178, 409

\bibitem[{{Mendez} \& {Niemela}(1981)}]{MN81}
---. 1981, ApJ, 250, 240

\bibitem [{{Moe} \& {De Marco}(2006)}]{MD06}
{Moe}, M. \& {De Marco}, O. 2006, AJ, in press

\bibitem[{{Napiwotzki}(1999)}]{Nap99}
{Napiwotzki}, R. 1999, A\&A, 350, 101

\bibitem[{{Napiwotzki} {et~al.}(1994){Napiwotzki}, {Heber}, \&
  {Koeppen}}]{Nap+94}
{Napiwotzki}, R., {Heber}, U., \& {Koeppen}, J. 1994, A\&A, 292, 239

\bibitem[{{Noskova}(1980)}]{Nos80}
{Noskova}, R.~I. 1980, Astronomicheskij Tsirkulyar, 1128, 1

\bibitem[{{Pollacco} \& {Bell}(1993)}]{PB93}
{Pollacco}, D.~L. \& {Bell}, S.~A. 1993, MNRAS, 262, 377

\bibitem[{{Pollacco} \& {Bell}(1994)}]{PB94}
---. 1994, MNRAS, 267, 452

\bibitem[{{Sahai} \& {Trauger}(1998)}]{ST98}
{Sahai}, R. \& {Trauger}, J.~T. 1998, AJ, 116, 1357

\bibitem[{{Soker}(2006)}]{Sok06}
{Soker}, N. 2006, PASP, 118, 260

\bibitem[{{Soker} \& {Subag}(2005)}]{SS05}
{Soker}, N. \& {Subag}, E. 2005, AJ, 130, 2717

\bibitem[{{Sorensen} \& {Pollacco}(2003)}]{SP03}
{Sorensen}, P.~M. \& {Pollacco}, D.~L. 2003, in Astronomical Society of the
  Pacific Conference Series, ed. R.~L.~M. {Corradi}, J.~{Mikolajewska}, \&
  T.~J. {Mahoney}, 494--+

\bibitem[{{Strassmeier} {et~al.}(1997){Strassmeier}, {Hubl}, \&
  {Rice}}]{Str+97}
{Strassmeier}, K.~G., {Hubl}, B., \& {Rice}, J.~B. 1997, A\&A, 322, 511

\bibitem[{{Tovmassian} {et~al.}(2004){Tovmassian}, {Napiwotzki}, {Richer},
  {Stasi{\'n}ska}, {Fullerton}, \& {Rauch}}]{Tov04}
{Tovmassian}, G.~H., {Napiwotzki}, R., {Richer}, M.~G., {Stasi{\'n}ska}, G.,
  {Fullerton}, A.~W., \& {Rauch}, T. 2004, ApJ, 616, 485

\bibitem[{{Tylenda} {et~al.}(1989){Tylenda}, {Acker}, {Gleizes}, \&
  {Stenholm}}]{Tyl+89}
{Tylenda}, R., {Acker}, A., {Gleizes}, F., \& {Stenholm}, B. 1989, A\&AS, 77,
  39

\bibitem[{{van Winckel}(2003)}]{van03}
{van Winckel}, H. 2003, in Astronomical Society of the Pacific Conference
  Series, ed. R.~L.~M. {Corradi}, J.~{Mikolajewska}, \& T.~J. {Mahoney}, 294--+

\bibitem[{{Walsh} \& {Walton}(1996)}]{WW96}
{Walsh}, J.~R. \& {Walton}, N.~A. 1996, A\&A, 315, 253

\bibitem[{{West}(1976)}]{Wes76}
{West}, R.~M. 1976, PASP, 88, 896

\bibitem[{{Wood} {et~al.}(1995){Wood}, {Robinson}, \& {Zhang}}]{Woo+95}
{Wood}, J.~H., {Robinson}, E.~L., \& {Zhang}, E.-H. 1995, MNRAS, 277, 87

\end{thebibliography}

\end{document}